\documentclass[aps,prd,twocolumn,superscriptaddress]{revtex4}
\usepackage{graphicx}% Include figure files
\usepackage{dcolumn}% Align table columns on decimal point
\usepackage{bm}% bold math
\usepackage{subfigure}
\usepackage{amssymb}
\usepackage{amsmath}
\usepackage{braket}
\usepackage{caption}
\newcommand{\beq}{\begin{equation}}
\newcommand{\eeq}{\end{equation}}
\newcommand{\beqar}{\begin{eqnarray}}
\newcommand{\eeqar}{\end{eqnarray}}
\newcommand{\ds}{\displaystyle}
\newcommand{\rmd}{{\rm d}}
\newcommand{\Tr}{{\rm Tr}}
\newcommand{\const}{{\rm const}}
\usepackage{color}

\begin{document}
%\begin{frontmatter}
%\draft

\title{Unruh effect and information entropy approach}

\author{M.~V.~Teslyk}
\affiliation{
Taras Shevchenko National University of Kyiv, UA-01033 Kyiv, Ukraine
\vspace*{1ex}}
\author{O.~M.~Teslyk}
\affiliation{
Taras Shevchenko National University of Kyiv, UA-01033 Kyiv, Ukraine
\vspace*{1ex}}
\author{L.~V.~Zadorozhna}
\affiliation{
Taras Shevchenko National University of Kyiv, UA-01033 Kyiv, Ukraine
\vspace*{1ex}}
\author{L.~Bravina}
\affiliation{
Department of Physics, University of Oslo, PB 1048 Blindern,
N-0316 Oslo, Norway
\vspace*{1ex}}
\author{E.~Zabrodin}
\altaffiliation[Also at ]{
Skobeltsyn Institute of Nuclear Physics,
Moscow State University, RU-119991 Moscow, Russia
\vspace*{1ex}}
\affiliation{
Department of Physics, University of Oslo, PB 1048 Blindern,
N-0316 Oslo, Norway
\vspace*{1ex}}

\date{\today}

\begin{abstract}
Total entropy generated by the Unruh effect is calculated within the
framework of information theory. In contrast to previous studies, here
the calculations are done for the finite time of existence of the 
non-inertial reference frame. In this case only the finite number of 
particles is produced. Dependence on mass of the emitted particles is 
taken into account. 
Analytic expression for the entropy of radiated boson and fermion 
spectra is derived. We study also its asymptotics corresponding to
limiting cases of low and high acceleration. 
The obtained results can be further generalized to other intrinsic 
degrees of freedom of the emitted particles, such as spin and electric 
charge.
\end{abstract}

\maketitle

\section{Introduction}
\label{sec:intro}

As was demonstrated by Unruh \cite{u_Unruh}, the observer comoving with
non-inertial reference frame (RF) with acceleration $a$ will detect 
particles thermalized at temperature $$T = \frac{a}{2\pi}$$ in Planck 
units, whereas the observer in any inertial RF will see bare vacuum. 
If acceleration $a$ equals to surface gravity of some Schwarzschild 
black hole (BH), then $T$ coincides with the temperature $T_\mathrm{BH}$ 
of Bekenstein-Hawking radiation \cite{Haw_75,Bek_73} of the horizon.

This peculiar non-invariance of vacuum has raised a lot of interest to 
the topic, for review see, e.g., \cite{RevModPhys.80.787} and references
therein. Recall that the Unruh effect initially was derived for scalar 
particles. Here the change in the ratio between negative and positive 
frequency modes of scalar fields in the noninertial RF was considered 
\cite{u_Unruh}. 
Generalizations to arbitrary trajectories of the observer is 
discussed in \cite{PhysRevD.75.065006}, whereas the generalization to
accelerated reference frames with rotation can be found in \cite{u_rot}.
The emergence of the Unruh effect in Rindler manifold of arbitrary
dimension and its relationship to the vacuum noise and stress are 
investigated in \cite{Ta_86}. 
Various methods and approaches have been employed. For instance, 
algebraic approach was used to extend the Unruh effect to theories
with arbitrary spin and with interaction \cite{BiWi_75}, whereas the
path integral approach was applied to derive the effect for fermions
within the framework of quantum field theory \cite{UnWe_84}. 
Among the recent studies one can mention the relativistic quantum
statistical mechanics approach \cite{Be_18,BeRi_19,PTZ_19} based on 
application of Zubarev's density operator \cite{ZPS_79,BBG_19}. Within 
this approach the Unruh effect was obtained first for scalar particles
\cite{Be_18} and then generalized to the gas of massless fermions
\cite{PTZ_19}. In the present study we employ the approach based on
application of the information theory which was not elaborated
extensively yet.
 
Usually the non-inertial observer is assumed to accelerate forever. 
However, such an assumption implies availability of infinite energy 
supply and ever-lasting particle emission. The more sophisticated 
scenario which considers the Unruh effect at finite time interval is 
analyzed in \cite{PhysRevD.46.5267,cqg_padmanabhan_1996}.

There is a lot of proposals for the detection and application of the 
Unruh effect, see, e.g., \cite{ChTa_99,LoSa_08,Akh_08}. 
Paper \cite{u_crypt} discusses possibility of eavesdropping in the 
non-inertial reference frame. Production of the entangled photon pairs 
from the vacuum with the help of the Unruh effect was investigated in 
\cite{u_photon}, whereas in \cite{PaMa_97} creation of accelerated
black holes by means of the Unruh effect was studied. 
In \cite{BeLe_83} the authors 
discuss possibility of using accelerated electrons as thermometers. 
Generated bosons and fermions were considered as being produced via 
the quantum tunneling mechanism at the Unruh horizon in 
\cite{0903.0250,0908.3149}.

The Unruh effect may be considered as a source of particle 
production. The idea has been widely employed 
\cite{BaTr_78,GrSr_79,Ho_79,BBGM_80,KhTu_05,CKhS_07,CaSa_14,BGS_12} 
in order to explain 
multiparticle production in hadronic and heavy-ion collisions at
ultrarelativistic energies. The attractive feature of application of
the Unruh effect as possible mechanism of the multiparticle production
is the thermalized spectra of newly produced particles. Experiments 
with ultrarelativistic hadronic and heavy-ion collisions and their
theoretical interpretations indicate that the produced matter seems
to reach equilibrium extremely quickly, see, e.g., \cite{qm19,qm18}
for present status of the field. The mechanism of this fast
equilibration is still debated, therefore, the Unruh effect might
be of a great help.
At the same time, since the Unruh source is thermal, it results in 
observer-dependent entropy generation \cite{MMR_04}. In the present
paper we also consider the Unruh horizon as a thermal source of 
particles. These particles are characterized by thermal distribution. 
Our aim is to estimate the entropy of the distribution and to 
define its dependence on any intrinsic degrees of freedom of the 
emitted particles.

The paper is organized as follows. Section~\ref{sec:2} presents the
necessary basics from probability theory and information theory. 
Section~\ref{sec:3} briefly describes Unruh effect and density matrix 
of the emitted quanta. The total entropy of the Unruh source is 
estimated in Sect.~\ref{sec:4}. Here the general expression for the 
entropy of fermion and boson radiation is derived, as well as its 
analytic series expansion. In Sect.~\ref{sec:5} one is dealing with 
the analysis of temperature asymptotics of the entropy. Two limiting
cases corresponding to low and high temperature, or, equivalently, 
acceleration of the observer, are considered. Section~\ref{sec:6} 
is devoted to contribution of intrinsic degrees of freedom of the 
produced particles. Final remarks and conclusions can be found in 
Sect.~\ref{sec:conclusions}.

\section{Probability and entropy}
\label{sec:2}

Let us consider some distribution $ \{X\} $ with unnormalized 
distribution probability $ d\left(x\right) $. In other words, 
$ d\left(x\right) $ is a number of events in which $ x $ has being 
observed. Shannon entropy $ H\left(X\right) $ for it may be written as
\begin{equation}
\ds
\label{H(X)}
\begin{split}
	H\left(X\right)
	       &= -\sum_{x}
			\frac{d\left(x\right)}{\mathcal{D}_{X}}
			  \ln\frac{d\left(x\right)}{\mathcal{D}_{X}} \\
	       &= \ln\mathcal{D}_{X}
		  - \frac{1}{\mathcal{D}_{X}}
		    \sum_{x}
		         d\left(x\right)
		     	  \ln d\left(x\right) \ ,
\end{split}
\end{equation}
where $\mathcal{D}_{X} = \sum_{x} d\left(x\right)$. $H\left(X\right)$ 
encodes amount of information we need in order to completely describe 
$ \{X\} $, i.e., this is amount of information we are lacking. Therefore, 
we should deal with the distribution $ \{X\} $. It is scale-invariant, 
so it does not changed under the transformations $ d\left(x\right) \to 
\alpha d\left(x\right) $ for any $ \alpha = const$.

Similarly, for joint distribution $ \{X,Y\} $ with unnormalized 
distribution probability $ d\left(x,y\right) $ one can write down 
Shannon entropy $ H\left(X,Y\right) $ as
\begin{equation}
\ds
\label{H(X,Y)}
\begin{split}
	H\left(X,Y\right)
		&= -\sum_{x,y}
		  \frac{d\left(x,y\right)}{\mathcal{D}_{X,Y}}
	          \ln\frac{d\left(x,y\right)}{\mathcal{D}_{X,Y}} \\
		&= \ln\mathcal{D}_{X,Y}
		 - \frac{1}{\mathcal{D}_{X,Y}}
		   \sum_{x,y}
		  		d\left(x,y\right)
		  		\ln d\left(x,y\right),
\end{split}
\end{equation}
where $ \mathcal{D}_{X,Y} = \sum_{x,y} d\left(x,y\right) $.

In the joint case one may define conditional probability 
$ d\left(x|y\right) $ as
\begin{equation}
\ds
\label{d(x|y)}
	d\left(x|y\right)
		= \frac{d\left(x,y\right)}{d\left(y\right)}\ , \qquad
	d\left(y\right)
		= \sum_x d\left(x,y\right).
\end{equation}
It defines the amount of events with $ x $ from the set of events in 
which $ y $ occurs. Using Eq.~\eqref{H(X)}, Shannon entropy 
$ H\left(X|y\right) $ becomes
\begin{equation}
\ds
\label{H(X|y)}
\begin{split}
	H\left(X|y\right)
		&= \ln\mathcal{D}_{X|y}
		 - \frac{1}{\mathcal{D}_{X|y}}
		   \sum_{x}
		  		d\left(x|y\right)
		  		\ln d\left(x|y\right)\\
		&= -\sum_{x}
				d\left(x|y\right)
				\ln d\left(x|y\right) ,
\end{split}
\end{equation}
where $ \mathcal{D}_{X|y} = \sum_x d\left(x|y\right) = 1 $, as follows
from Eq.~\eqref{d(x|y)}.

Finally, substituting Eq.~\eqref{d(x|y)} and Eq.~\eqref{H(X|y)} into 
Eq.~\eqref{H(X,Y)} one gets
\begin{equation}
\ds
\label{H conditional}
\begin{split}
	H\left(X,Y\right)
	       &= H\left(Y\right)
		+ \Braket{H\left(X\vert y\right)}_{Y} \\
	       &= H\left(X\right)
		+ \Braket{H\left(Y\vert x\right)}_{X} ,
\end{split}
\end{equation}
where averaging taken over ${X}$ or ${Y}$ reads
\begin{equation*}
	\Braket{\mathcal{A}}_{Z}
		= \frac{1}{\mathcal{D}_Z}
		  \sum_{z}
		  		d\left(z\right)\mathcal{A}\ ,\qquad
	Z \equiv X, Y.
\end{equation*} 

Recall, that all the formulae above are valid for discrete distributions 
only. In the continuous case one should use probability density function 
(PDF) $ p\left(x\right) $ instead of $ d\left(x\right) $. Shannon entropy 
becomes dimensional incorrect then and should be re-defined, as shown
in \cite{jaynes1,jaynes2}.

For distribution $\{X\} $ with PDF $p\left(x\right)$ the entropy 
given by Eq.~\eqref{H(X)} is generalized to
\begin{equation}
\ds
\label{H(X_p)}
	H\left(X_p\right)
		= \ln\mathcal{D}_{X_p}
		- \frac{1}{\mathcal{D}_{X_p}}
		  \int
		  	p\left(x\right)
		  	\ln p\left(x\right)
		  	\rmd x
		- \Braket{\ln\rmd x}_{X_p},
\end{equation}
where $ \mathcal{D}_{X_p} = \int p\left(x\right)\rmd x $ is the norm and
\begin{equation*}
	\Braket{\mathcal{A}}_{X_p}
		= \frac{1}{\mathcal{D}_{X_p}}
		  \int
		  	p\left(x\right)
		  	\mathcal{A}
		  	\rmd x.
\end{equation*}
The last term in Eq.~\eqref{H(X_p)} is related to the limiting density 
of discrete points and takes into account amount of information encoding 
discrete-continuum transition, see \cite{jaynes1,jaynes2} for details.
Note, that one may formally reduce $H\left(X_p\right)$ to 
$H\left(X\right)$ by substituting $ \int p\left(x\right)\rmd x$ into 
$\sum_{x}d\left(x\right)$ and setting $ \left\langle\ln\rmd x\right
\rangle_{X_p} $ to zero; the same procedure is valid also in the
opposite direction.

\section{Unruh effect}
\label{sec:3}

From here we will use Planck (or natural) units, $c = G = \hbar = k_B 
= 1$. Also, we restrict our analysis to 
$ 1+1 $-dimensional space-time, because two other spatial dimensions play 
no role and, therefore, can be neglected.

As was already mentioned in Sec.~\ref{sec:intro}, vacuum is non-invariant 
with respect to the reference frame \cite{u_Unruh}. In the non-inertial 
RF determined with the acceleration $ a $ one meets with the appearance 
of horizon separating space-time to the inside and outside domains. As a 
result, the non-inertial observer detects radiation going out from the 
horizon, while the inertial one detects the vacuum state $ |0\rangle $ 
only. For bosons the latter reads 
\cite{RevModPhys.80.787,0903.0250,0908.3149}
\begin{equation}
\ds
\label{|0> boson}
	\Ket{0}
	= \sqrt{\frac{1-\exp{\left(-E/T\right)}}{1-\exp{\left(-NE/T\right)}}}
		  \sum_{n=0}^{N-1}
		  		\exp{\left(-nE/2T\right)}
		  		\Ket{n}_{\rm in}
		  		\Ket{n}_{\rm out} \ ,
\end{equation}
whereas for fermions one gets
\begin{equation}
\ds
\label{|0> fermion}
	\Ket{0}
		= \frac{1}{\sqrt{1+\exp{\left(-E/T\right)}}}
		  \sum_{n=0}^{1}
		  \exp{\left(-nE/2T\right)}
		  \Ket{n}_{\rm in}
		  \Ket{n}_{\rm out}
\end{equation}
Here $ E $ is the energy of the quanta emitted at Unruh horizon with 
temperature $ T = a/\left(2\pi\right) $. Parameter $ N $, as can be seen 
from Eq.~\eqref{|0> boson}, encodes maximum amount of quanta at energy 
$ E $ plus 1. Loosely speaking $ N $ is a number of dimensions of the 
corresponding Fock space at given energy $E$ and temperature $T$ of the 
source. The subscripts {\it in} and {\it out} denote the components of 
the field with respect to the horizon. 

Usually $N$ is assumed to be infinite. But taking $ N \to \infty $ in 
Eq.~\eqref{|0> boson}, as it is widely used in the literature on the 
topic, seems to be too strong assumption, because the source is 
considered to produce infinite amount of energy, $ \left(N - 1\right)E 
\to \infty $. This is valid in case of everlasting acceleration that 
can be provided with the infinite energy supply only. Such a case seems
to be rather unlikely, therefore we assume maximum number of particles 
to be finite in all the calculations below. Also, let us consider only
boson production in what follows, because the expression for fermions
given by Eq.~\eqref{|0> fermion} can be derived from 
Eq.~\eqref{|0> boson} by setting $N = 2$.

Expression \eqref{|0> boson} is Schmidt decomposition 
\cite{a_pathak_13}. The outgoing radiation is described by density 
matrix
\begin{equation}
\ds
\label{rho}
\begin{split}
	\rho_{\rm out}
	       &= \Tr_{\rm in}
				\Ket{0}\Bra{0} \\
	       &= \frac{1-\exp{\left(-E/T\right)}}
                       {1-\exp{\left(-NE/T\right)}}
		  \sum_{n=0}^{N-1}
		  	\exp{\left(-\frac{nE}{T}\right)}
		  	\Ket{n}_{\rm out}\Bra{n}_{\rm out}\ ,
\end{split}
\end{equation}
where we have traced over the inaccessible degrees of freedom 
({\it in-} modes). Thus, pure vacuum state from the inertial RF has 
transformed into the mixed one in the non-inertial RF. Here appears
the geometric origin of the Unruh effect. Namely, finiteness of the 
speed of light leads to appearance of the horizon dividing the all 
modes in Hilbert space into the accessible ({\it out-}) and 
non-accessible ({\it in-}) ones. The complete state is obviously pure 
and follows unitary evolution. But because one has limited access to 
it in the non-inertial RF, it looks like a decoherence.
Eigenvalues of density matrix $\rho_{\rm out}$ define emission 
probability of certain number of particles at energy $E$ and 
temperature $T$. Therefore, Eq.~\eqref{rho} describes conditional 
multiplicity distribution $\{n|N,E,T\}$ at given $N,\ E$ and $T$.

\section{Unruh entropy}
\label{sec:4}

For the emission probability $\rho_{\rm out}$ from Eq.~\eqref{rho} 
the von Neumann entropy is defined as
\begin{equation}
\ds
\label{H(n|N,E,T)}
\begin{split}
	H\left(\rho_{\rm out}\right)
	       &= -\Tr \rho_{\rm out} \ln\rho_{\rm out}
		= H\left(n|N, E,T\right) \\
	       &= \sigma\left(qE/T\right)\Big|_{q=N}^{q=1}\ ,
\end{split}
\end{equation}
where we use the following notations
\begin{equation}
	\sigma\left(qE/T\right)
		= \frac{qE/T}{\exp{\left(qE/T\right)}-1}
		 - \ln{\left[1-\exp{\left(-\frac{qE}{T}\right)}\right]}\ , 
\end{equation}
\begin{equation}
	f\left(x\right)\Big|^{x=a}_{x=b}
		= f\left(a\right)
		 - f\left(b\right)\ .
\end{equation}

As one may notice, $ H\left(n|N, E/T\right) $ is an even function of 
$E/T$, i.e. $ H\left(n|N, E/T\right) = H\left(n|N, -E/T\right) $. 
Asymptotic behavior of entropy \eqref{H(n|N,E,T)} with respect to $E/T$ 
is the following
\begin{equation}
\ds
\label{H(n|N,E/T) limits}
\begin{split}
	\lim_{E/T\to 0}H\left(n|N, E,T\right)
		&= \ln N
		 = \max\left(H\right)\\
	\lim_{E/T\to\infty}H\left(n|N, E,T\right)
		&= 0\ .
\end{split}
\end{equation}

Expression \eqref{H(n|N,E,T)} defines entropy of the emitted quanta, as 
well as the quanta inside the horizon, for some mode of the radiated 
field only, which is determined by parameter $N$, energy $E$ and 
temperature $T$.
Parameter $N$ depends on amount of time during which the observer is 
being described by non-inertial reference frame. It follows from the 
fact that the longer one is observing the horizon, the more particles 
at any fixed energy may be detected. Therefore we conclude that $N$ 
should increase with time.
Temperature $T$ is completely determined by the acceleration $a$, 
see \cite{u_Unruh}.
However, $E$ cannot be considered as a fixed parameter. The 
non-inertial observer is expected to detect particles at different 
energies. Energy range for the particles may be written as
\begin{equation*}
	m \leq E \leq M\ ,
\end{equation*}
where $m$ is invariant mass of the particles, and $M$ is the 
maximum energy to be observed, respectively. We assume $M$ to be 
limited by acceleration $a$, since observation of the high energy 
particles is very unlikely due to energy conservation law: one 
cannot extract more energy from the vacuum than is being spent to 
sustain the observer's acceleration.

Unfortunately, definition of the energy range does not mean we 
know the spectrum distribution $\{E\}$. It is determined by 
unnormalized PDF $ p\left(E\right)$ of emission of a particle from 
vacuum at energy $E$.

In order to figure out $p\left(E\right)$ somehow we use the 
following procedure. As can be noticed from Eq.~\eqref{rho}, for any 
particle number $n > 0$ the emission probability is proportional to 
factor $\exp{(-E/T)}$. Case with $n = 0$ means no emission at all. 
Therefore, one should expect exponential behavior for $p\left(E\right)$ 
\begin{equation}
\ds
\label{p(E)}
	p\left(E\right) = C \exp{\left(-E/T\right)}\ ,
\end{equation}
where prefactor $C$ is responsible for any corrections that might 
depend on the particle type and its quantum numbers. For the sake of
simplicity we assume $ C = \const $ and, therefore, drop it due to 
normalization reasons, see Sect.~\ref{sec:2}, in what follows. It is
worth noting that such assumption results in Schwinger-like mechanism 
of particle production \cite{Sch_51}. Thus we recovered Schwinger-like 
particle production from the properties of Hilbert space and space-time 
only. Recall, however, that this result is generated by 
Unruh effect after neglecting all the possible corrections.

Now we have spectrum distribution $ \{E\} $ as given by 
Eq.~\eqref{p(E)}. Without any loss of generality we assume energy to 
be defined within the range $ m\leq E\leq M $. From 
Eq.~\eqref{H conditional} and Eq.~\eqref{H(X_p)} one gets
\begin{widetext}
\begin{equation}
\ds
\label{H(n,E|N,T) before integration}
%\begin{split}
	H\left(n,E|N,T\right)
		= -\Braket{\ln\rmd E}_{E_p}
		 + \ln\mathcal{D}_{E_{p}}
		 - \frac{1}{\mathcal{D}_{E_{p}}}
		   \int_{m}^{M}
		   		p\left(E\right)
		   		\ln p\left(E\right)
		   		\rmd E
%		 &\quad
		  + \frac{1}{\mathcal{D}_{E_p}}
		    \int_{m}^{M}
		 		p\left(E\right)
		 		H\left(n|N,E,T\right)
		 		\rmd E.
%\end{split}
\end{equation}
In order to get analytic expression, we substitute Eq.~\eqref{p(E)} 
and Eq.~\eqref{H(n|N,E,T)} into 
Eq.~\eqref{H(n,E|N,T) before integration} and obtain after the 
straightforward calculations total Unruh entropy 
$H\left(n,E|N,T\right)$ in a form
\begin{equation}
\ds
\label{H(n,E|N,T)}
\begin{split}
	H\left(n,E|N,T\right)
		= &-\Braket{\ln\rmd E}_{E_p}
		 + 1
		 + \ln\mathcal{D}_{E_{p}}
		 + \frac{m \exp{\left(-m/T\right)} - 
                   M\exp{\left(-M/T\right)}}{\mathcal{D}_{E_{p}}}\\
%	\quad
		  &+ \frac{T}{\mathcal{D}_{E_{p}}}
		   \sum_{k=1}^{\infty}
		   	\Bigg\{
		   		\left[
		   			\frac{2kq+1}{k\left(kq+1\right)}
		   				+ q\frac{E}{T}
		   			\right]
%		&\hspace{7em}
                       \times
		   		\frac{\exp{\left[-(kq+1)E/T\right]}}{kq+1}
				\bigg|_{E=M}^{E=m}
			\Bigg\}
			\Bigg|_{q=N}^{q=1}\ ,
\end{split}
\end{equation}
\end{widetext}
where
\begin{equation}
\ds
\label{p(E) norm}
  \mathcal{D}_{E_{p}}
       = \int_{m}^{M}
         p\left(E\right) \rmd E
       = T\left[\exp{\left(-\frac{m}{T}\right)} - 
                \exp{\left(-\frac{M}{T}\right)}\right]
\end{equation}
and $ \sigma\left(qE/T\right) $ from Eq.~\eqref{H(n|N,E,T)} is 
represented by the following series
\begin{equation}
\ds
	\sigma(qE/T)
		= \sum_{k=1}^{\infty}
			\left(
				  \frac{1}{k}
				+ q\frac{E}{T}
			\right)
			\exp{\left(-\frac{kqE}{T}\right)}.
\end{equation}

The first term in Eq.~\eqref{H(n,E|N,T)} is responsible for encoding 
discrete-continuum transition, see \cite{jaynes1,jaynes2}. It is 
expected to depend neither on any quantum numbers of outgoing particles 
nor on the reference frame. Therefore, we assume 
$ \Braket{\ln\rmd E}_{E_p} $ to be constant.

\begin{figure}
\resizebox{\linewidth}{!}{
             \includegraphics[scale=0.50]{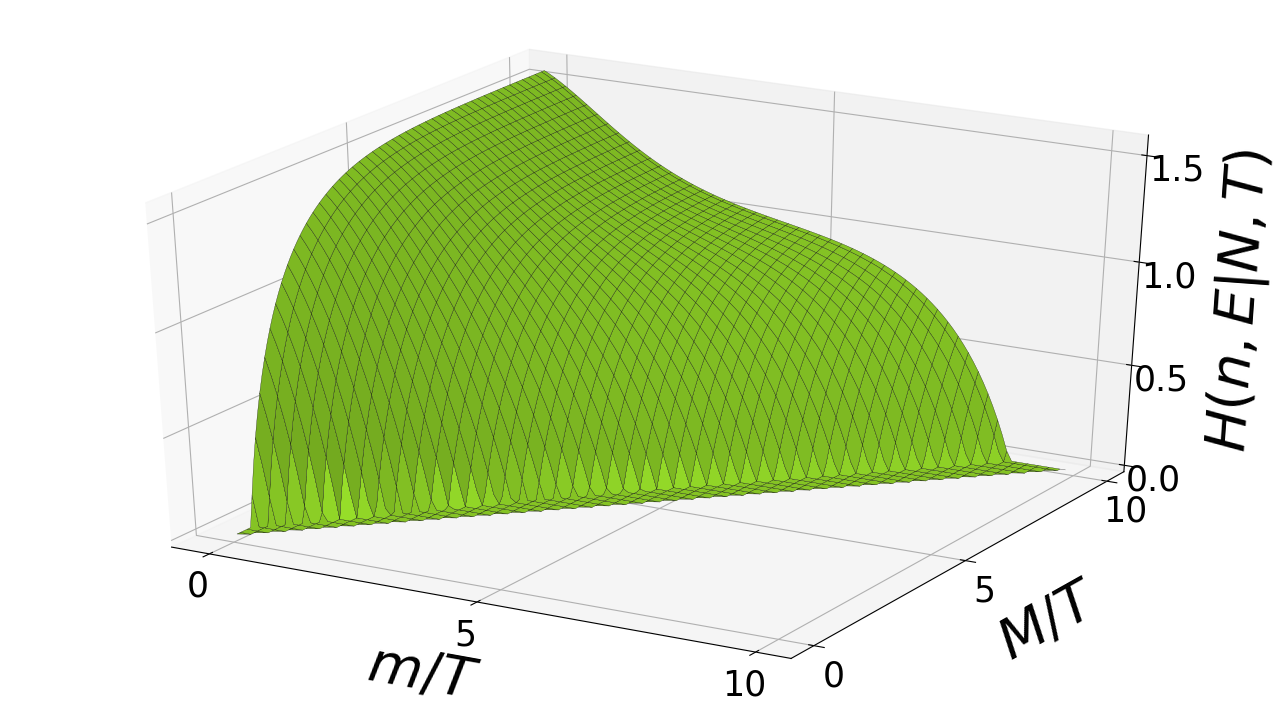}
}
\caption{(Color online)
Entropy $H(n,E|N,T)$ of Unruh radiation given by Eq.~\eqref{H(n,E|N,T)} 
for fermions $(N = 2)$ as function of $m/T$ and $M/T$.
}
\label{fig:entr_N_2}
\end{figure}

\begin{figure}
\resizebox{\linewidth}{!}{
                \includegraphics[scale=0.50]{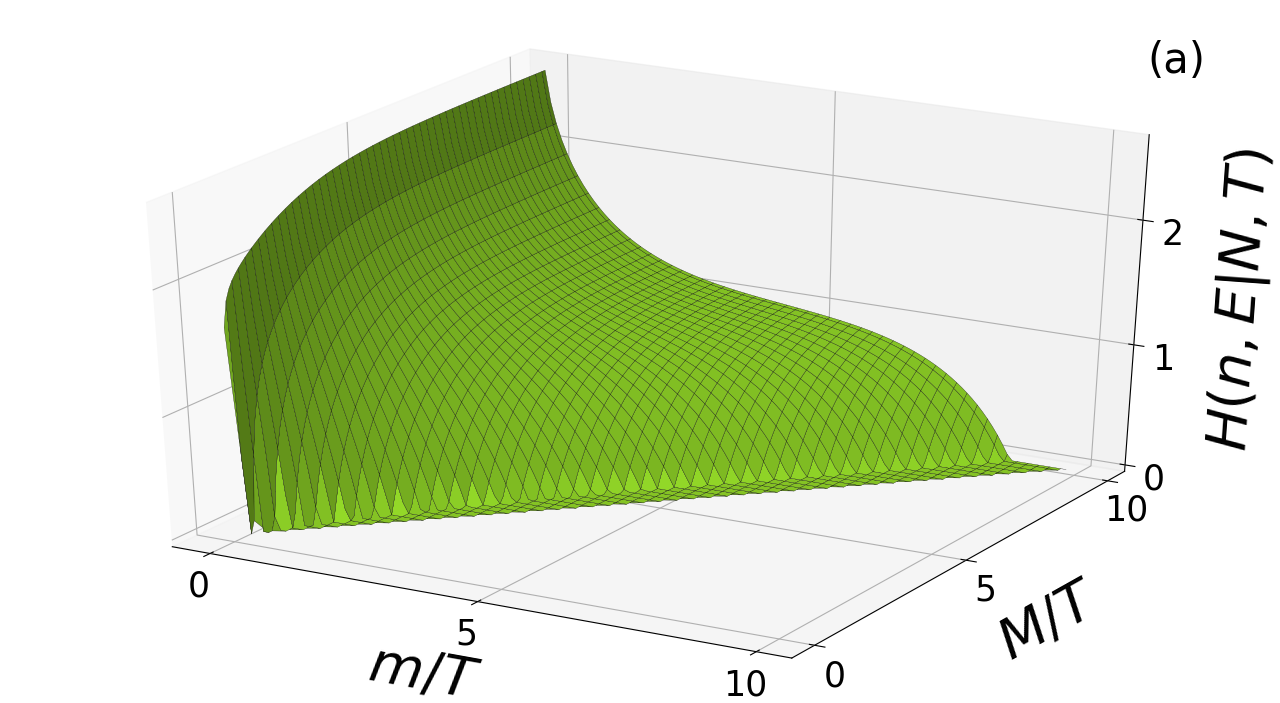}}
\resizebox{\linewidth}{!}{
                \includegraphics[scale=0.50]{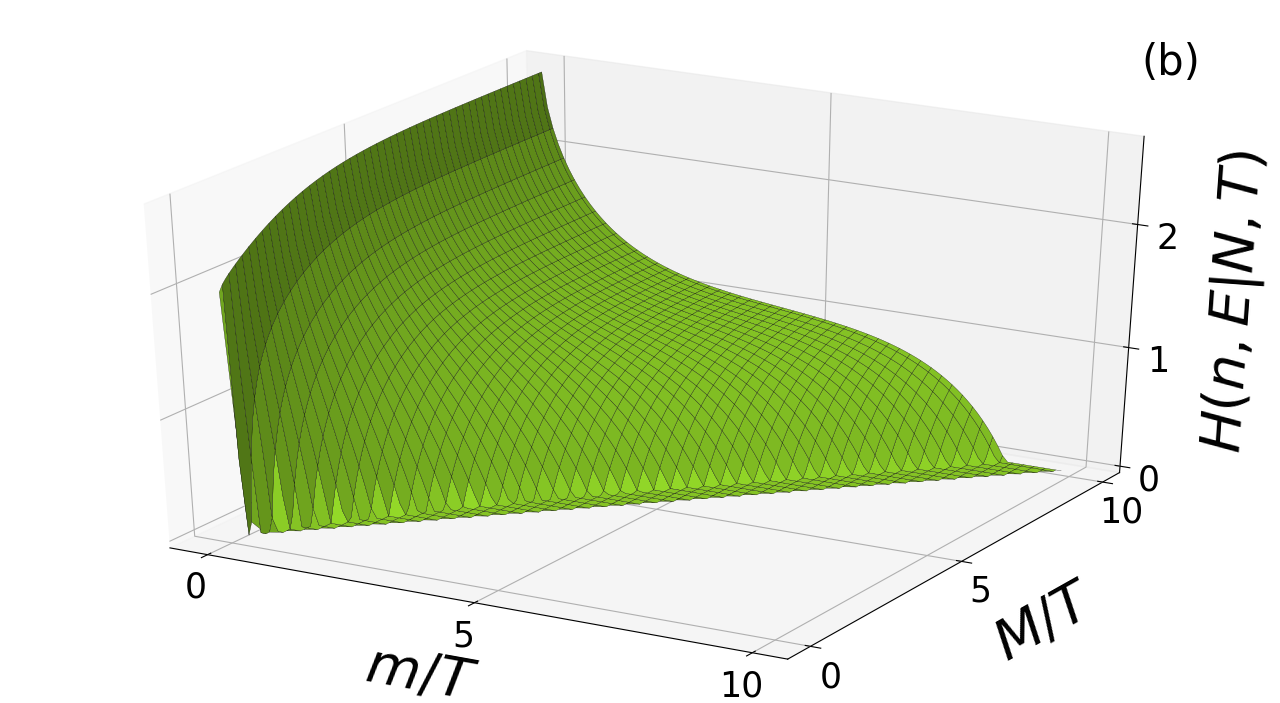}}
\caption{(Color online)
The same as Fig.~\ref{fig:entr_N_2} but for bosons. The spectrum of
bosons contains {\bf (a)} $N = 100$ and {\bf (b)} $N = 1000$ particles.
}
\label{fig:entr_N_100}
\end{figure}

Expression \eqref{H(n,E|N,T)} defines entropy for the distribution 
$ \{n,E|N,T\} $ of the particles being detected by the observer 
associated to non-inertial RF moving with acceleration $a = 2\pi T$. 
Recall, that in case of fermions one should use $ N = 2 $. For bosons 
$ N $ may take any positive integer value obeying the energy 
conservation law. The entropy calculated for the Unruh radiation of
fermions and bosons is presented in Fig.~\ref{fig:entr_N_2} and
Fig.~\ref{fig:entr_N_100}, respectively. One can see the distinct
maximum in the region of small values of $m/T$ ratio. The maximum
increases with rising $M/T$ ratio and becomes more pronounced with
the increase of radiated particles, see Fig.~\ref{fig:entr_N_100}.

The considered example seems to be straightforward. However, one should 
keep in mind that all the analysis above is valid for $1+1$-dimensional 
space-time. Other spatial dimensions do not contribute to the density 
matrix $ \rho_{\rm out} $ or to its von Neumann entropy, because the 
corresponding subspaces of the Hilbert space contribute to 
$ \rho_{\rm out} $ via direct tensor product and, therefore, can be 
traced out with no consequences to the analysis above.
This simple direct extension to additional spatial dimensions for Unruh 
effect may lead to widely-spread conclusion that Unruh effect results 
in appearance of thermal bath all over the space. In our opinion this 
conclusion needs to be clarified. Namely, in the last case the 
non-inertial observer, as well as the horizon itself, should be 
considered as an infinite plane in the additional spatial dimensions 
being accelerated alongside the normal to the plane. But the observer 
should be finite and, therefore, cannot detect particles from the 
half-space defined by the horizon. Otherwise, it would lead to 
faster-than-light speed communication and to causality violation, 
because the transition to inertial RF cannot cause immediate 
disappearance of the Unruh radiation from the horizon occupying the 
half-space.

To overcome the difficulties we have to assume that
\begin{itemize}
  \item In order to obey energy conservation law $ N $ should 
        be finite.
  \item In case of (2+1) or (3+1)-dimensional space-time the
Unruh horizon should be considered as radiation source of finite 
size. 
\end{itemize}

Due to axial symmetry of the non-inertial reference frame  the 
horizon should be of disk shape with some radius $ r $. The radius 
can be determined by the observer's size and causality, i.e. 
finiteness of light speed. Such an assumption leads to 
observer-dependent size of $ r $. The problem may be cured, e.g., if 
one considers the observer's acceleration $ a $ as a surface gravity 
of the corresponding black hole and obtain some efficient scale 
$ r = \left(4\pi T\right)^{-1} $.

One might be confused by the fact that since the Unruh effect describes
the thermal bath its entropy should be maximal. As can be easily 
noticed from the eigenvalues of the density matrix \eqref{rho}, all of 
them exponentially depend on the total energy of the emitted number of 
particles and thus generate a well-known partition function. Note,
however, that $ \rho_{\rm out} $ is defined for some \emph{fixed} value 
of energy. Therefore, $ E $ can be considered as a parameter of the 
conditional distribution $ \left\{n|N,E,T\right\} $. Dealing with joint 
distribution $ \left\{n,E|N,T\right\} $ over multiplicity $ n $ and 
energy $ E $ of the emitted quanta one should take into account energy 
conservation. It results in some correlations between the possible 
number of emitted particles and their energy. Thus, the entropy 
$ H\left(n,E|N,T\right) $ describes not a completely thermal source but 
some other one.

\section{Asymptotics of Unruh entropy}
\label{sec:5}

Let us analyze asymptotic behavior of total Unruh entropy 
\eqref{H(n,E|N,T) before integration} for (i) small and (ii) large 
acceleration of the observer.
The case of small acceleration is analogous to $ T \to 0 $, 
therefore, we will drop all but the leading term in 
Eq.~\eqref{H(n,E|N,T) before integration}.
At small temperatures Eq.~\eqref{p(E) norm} transforms to
\begin{equation}
\ds
\label{p(E) norm at T to 0 case B}
	\mathcal{D}_{E_p}\Big|_{T \to 0}
		\approx T \exp{\left(-m/T\right)},
\end{equation}
where we have neglected the term $\exp{(-M/T)}$ since $M$ is the 
upper bound for the energy spectrum and, therefore, $ M > m $.
The Unruh entropy becomes
\begin{equation}
\ds
\label{H(E) at T to 0 case B}
\begin{split}
	H\left(E\right)\Big|_{T \to 0}
		&= \ln\mathcal{D}_{E_p}
		 - \frac{1}{\mathcal{D}_{E_p}}
		   \int_{m}^{M}
				p\left(E\right)
				\ln p\left(E\right)
				\rmd E\\
		&\approx
		   \ln T
		 - \frac{m}{T}
		 + 1
		 + \frac{m}{T}
		 = \ln T + 1.
\end{split}
\end{equation}

\begin{figure}
\resizebox{\linewidth}{!}{
             \includegraphics[scale=0.60]{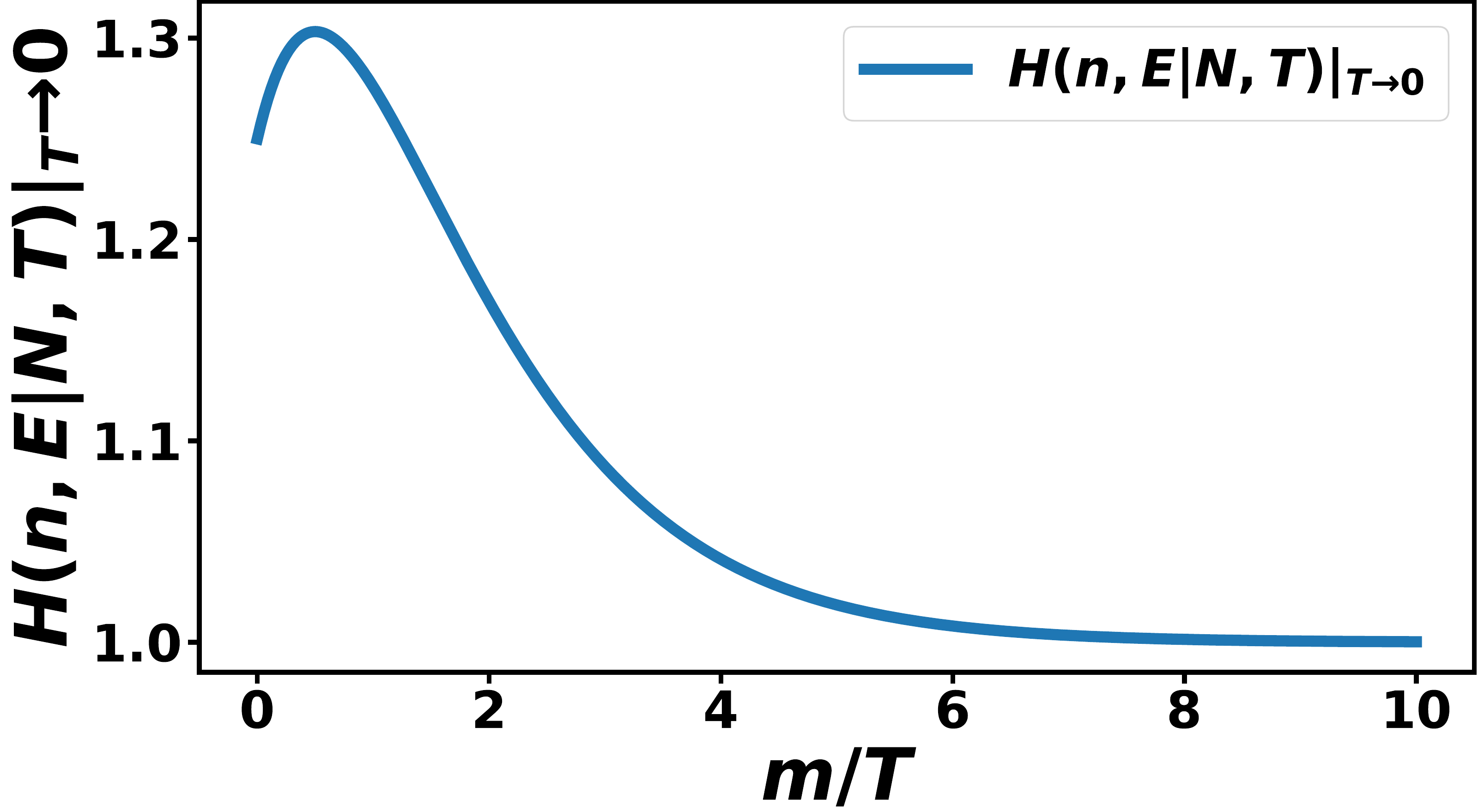}}
\caption{(Color online)
Asymptotic behavior of entropy $H(n,E|N,T)$ given by 
Eq.~\eqref{H(n,E|N,T) at T to 0 case B} at $T \rightarrow 0$ 
as function of $m/T$.
}
\label{fig:zero_asympt}
\end{figure}

Because the entropy $H\left(n|N,E,T\right)$ equals to zero when $N = 1$, 
we consider the case with $N > 1$ for $ T \to 0 $. Neglecting all higher 
order exponents, one obtains from Eq.~\eqref{H(n|N,E,T)} that
\begin{equation}
\ds
\label{H(n|N,E,T) at T to 0 case B}
	H\left(n|N,E,T\right)\Big|_{T \to 0}
		\approx \frac{E}{T}\exp{\left(-E/T\right)}.
\end{equation}

Substituting Eq.~\eqref{H(E) at T to 0 case B} and 
Eq.~\eqref{H(n|N,E,T) at T to 0 case B} into 
Eq.~\eqref{H(n,E|N,T) before integration} we get
\begin{widetext}
\begin{equation}
\ds
\label{H(n,E|N,T) at T to 0 case B}
%	T \to 0
%	\quad\Rightarrow\quad
	H\left(n,E|N,T\right)\Big|_{T \to 0}
		\approx   -\Braket{\ln\frac{\rmd E}{T}}_{E_p}
				+ 1
				+ \frac{1}{4}
				  \left(1 + \frac{2m}{T}\right)
				  \exp{\left(-m/T\right)},
\end{equation}
where all higher order exponents are omitted. This distribution is
displayed in Fig.~\ref{fig:zero_asympt}. The entropy reaches 
maximum at $m/T \approx 0.5$ and quickly drops to unity at larger
values of this ratio.

In case of large acceleration $a \to \infty \Leftrightarrow T \to 
\infty$ one obtains from Eq.~\eqref{p(E) norm}
\begin{equation}
\ds
\label{p(E) at T to infty}
%\begin{split}
	\int_{m}^{M}
		p\left(E\right)\Big|_{T \to \infty}
		\rmd E
			= \int_{m}^{M}
			   \left(
					1
				  - \frac{E}{T}
				  + \frac{E^2}{2T^2}
			   \right)
			   \rmd E
			 + \mathcal{O}\left(1/T^3\right)\\
			= \left(M - m\right)
			   \left(
			   		1
			   	  - \frac{M + m}{2T}
			   	  + \frac{M^2 + Mm + m^2}{6T^2}
			   \right)\\
%			&\quad
			 + \mathcal{O}\left(1/T^3\right)\ ,
%\end{split}
\end{equation}
and, therefore,
\begin{equation}\label{H(E) at T to infty}
%\begin{split}
%	T \to \infty
%	\quad\Rightarrow\quad
	H\left(E\right)
		= \ln{\mathcal{D}_{E_p}}
		 - \frac{1}{\mathcal{D}_{E_p}}
		   \int_{m}^{M}
		   		p\left(E\right)
		   		\ln{p\left(E\right)}
		   		\rmd E\\
		= \ln{\left(M - m\right)}
		 - \frac{\left(M - m\right)^2}{24T^2}
		 + \mathcal{O}\left(1/T^3\right)\ .
%\end{split}
\end{equation}
Thus the conditional entropy $ H\left(n|N,E,T\right) $ from 
Eq.~\eqref{H(n|N,E,T)} becomes
\begin{equation}
\ds
\label{H(n|N,E,T) at T to infty}
	H\left(n|N,E,T\right)\Big|_{T \to \infty}
		= \ln{N}
		- \frac{N^2 - 1}{24T^2}
		  E^2
		+ \mathcal{O}\left(1/T^4\right)\ ,
\end{equation}
that together with Eq.~\eqref{p(E) at T to infty} gives us
\begin{equation}
\ds
\label{<H(n|N,E,T)>-E-p at T to infty}
%\begin{split}
	\frac{1}{\mathcal{D}_{E_p}}
	\int_{m}^{M}
		p\left(E\right)
		H\left(n|N,E,T\right)
		\rmd E
			= \ln{N}
			 - \frac{M^2 + Mm + m^2}{72T^2}
			   \left(N^2 - 1\right)
			 + \mathcal{O}\left(1/T^3\right).
%\end{split}
\end{equation}

Finally, substituting Eq.~\eqref{H(E) at T to infty} and 
Eq.~\eqref{<H(n|N,E,T)>-E-p at T to infty} into 
Eq.~\eqref{H(n,E|N,T) before integration} we obtain the desired
asymptotics at high acceleration (or temperature)
\begin{equation}
\ds
\label{H(n,E|N,T) at T to infty case B}
%\begin{split}
	H\left(n,E|N,T\right)\Big|_{T \to \infty}
		= -\Braket{\ln \rmd E}_{E_p}
		 + 1
		 + \ln{\left(M - m\right)}
		 + \ln{N}
%	&\quad
		 - \frac{
		 			\left(N^2 + 2\right)
		 			\left(M^2 + m^2\right)
		 		  + \left(N^2 - 7\right)
		 		    Mm
	 		    }{72T^2}
% 		&\quad
		 + \mathcal{O}\left(1/T^3\right).
%\end{split}
\end{equation}
\end{widetext}

\begin{figure}
\resizebox{\linewidth}{!}{
             \includegraphics[scale=0.50]{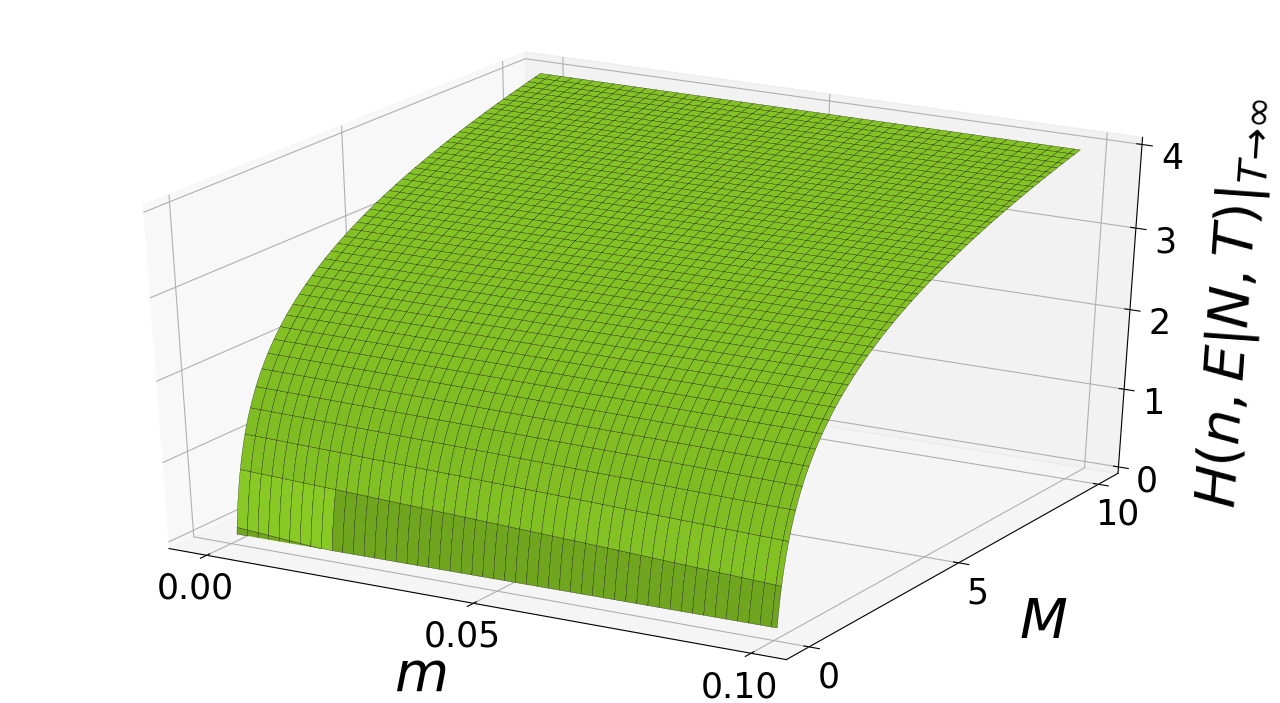}}
\caption{(Color online)
High-temperature asymptotics of the entropy $H(n,E|N,T)$ of Unruh 
radiation given by Eq.~\eqref{H(n,E|N,T) at T to infty case B}
for fermions $(N = 2)$ as function of $m$ and $M$.
}
\label{fig:infty_N_2}
\end{figure}

\begin{figure}
\resizebox{\linewidth}{!}{
                \includegraphics[scale=0.60]{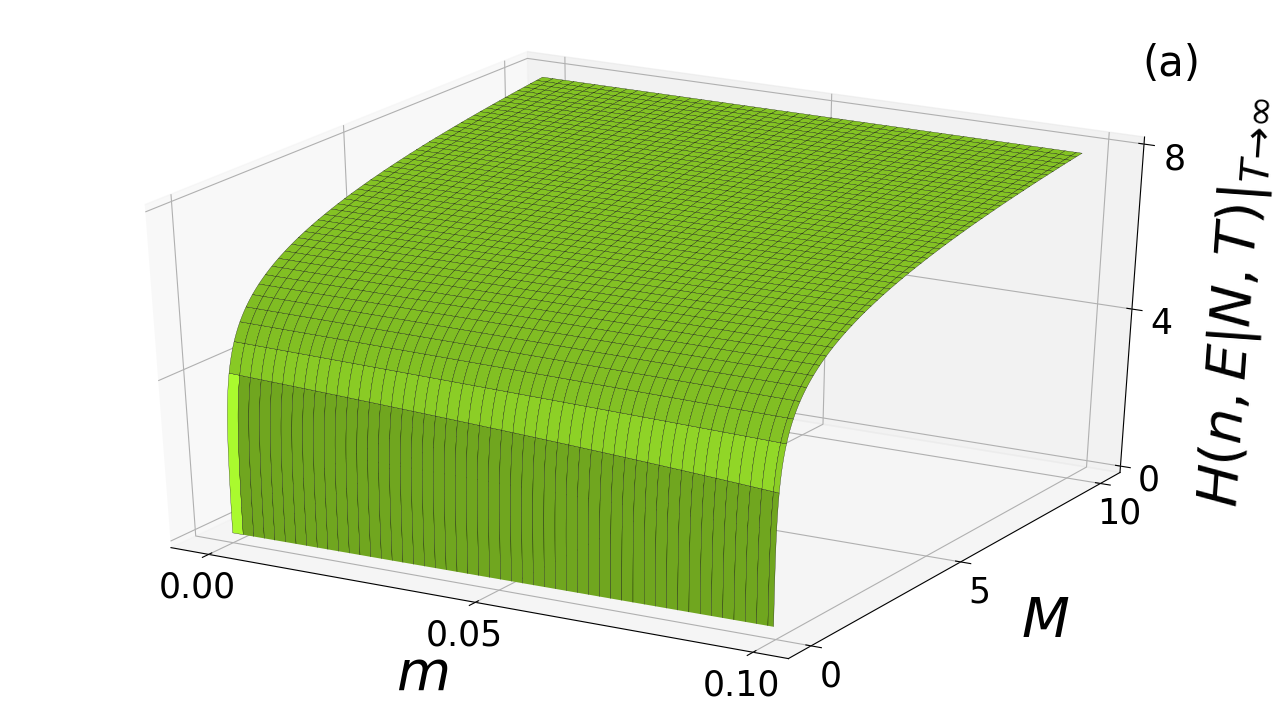}}
\resizebox{\linewidth}{!}{
                \includegraphics[scale=0.60]{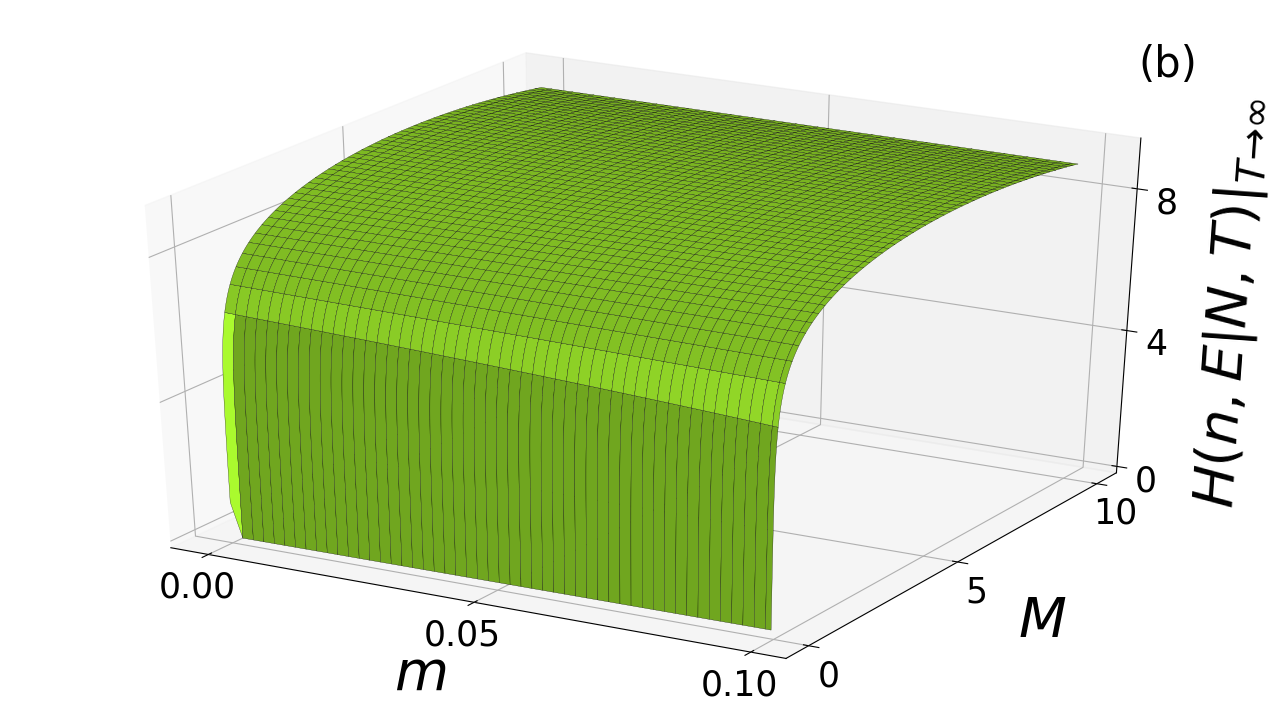}}
\caption{(Color online)
The same as Fig.~\ref{fig:infty_N_2} but for bosons with {\bf (a)} 
$N = 100$ and {\bf (b)} $N = 1000$ particles in the spectrum.
}
\label{fig:infty_N_100}
\end{figure}

The entropy asymptotics at $T \rightarrow \infty$ calculated according
to Eq.~\eqref{H(n,E|N,T) at T to infty case B} is presented in 
Fig.~\ref{fig:infty_N_2} for fermions $(N = 2)$ and in
Fig.~\ref{fig:infty_N_100} for boson spectra with $N = 100$ and 1000
particles, respectively. At high temperatures the entropy weakly depends
on $m$ and quickly increases with increasing value of $M$. The larger the
number of particles, the steeper the rising slope. For $N = 1000$ the 
entropy seems to saturate at $M \geq 5$. 

\section{Generalization to intrinsic degrees of freedom}
\label{sec:6}

Expression \eqref{H(n,E|N,T)} is valid for some mode of the radiated 
field only, which is defined by joint multiplicity-energy distribution 
$\left\{n,E\right\}$, temperature $T$ and parameter $N$. However, 
since the emitted particles may have additional degrees of freedom 
$\{\lambda\}$, such as electric charge, spin, polarization etc., they 
have to be taken into account too. This is equivalent to the following 
modification of the total distribution
\begin{equation*}
	\left\{n,E|N,T\right\} \to \left\{\lambda,n,E|N,T\right\}.
\end{equation*}
Using Eq.~\eqref{H conditional} we obtain then
\begin{equation}
\ds
\label{H(lambda,n,E|N,T)}
	H\left(\lambda,n,E|N,T\right)
		= H\left(\lambda\right)
		+ \Braket{H\left(n,E\vert N,T,\lambda\right)}_\lambda.
\end{equation}

But such a generalization is not an easy task at all.  Let us 
consider a simple example. While detecting particle at some $E$ one 
should measure its energy. Such a process results in consumption of 
the particle's momentum. One may argue that calorimetry is not 
required. The observer can build some source of similar particles and 
carry out interference experiments to determine energy of the particle 
to be detected. But any such interference will result in 
re-distribution of the momenta during interference, and therefore will 
change the observer's momentum as well. Thus one concludes that 
measuring the particle's energy $E$ leads to change of the observer's 
acceleration $a$. It implies change of Unruh temperature 
$T = a/\left(2\pi\right)$ of the source the observer is dealing with.

One may also note that Unruh effect is being considered in case of 
quasi-classical approach. It means that the density matrix 
$ \rho_{\rm out} $ in Eq.~\eqref{rho} is obtained under assumption 
that the outgoing radiation makes no influence on the background 
metric, see \cite{RevModPhys.80.787,0903.0250,0908.3149}.
Such a remark is correct, but what about other degrees of freedom 
$ \lambda $? For instance, taking into account spin of the particles 
emitted by the Unruh horizon may lead to change of the observer's 
angular momentum. In this case the observer's acceleration $a$ can 
not be constant due to conservation of the total angular momentum 
anyway and thus implies change of $T$ in Eq.~\eqref{H(lambda,n,E|N,T)} 
during particle identification.

So, the situation seems to be simple only if one neglects \emph{any}
influence of the outgoing particles during the Unruh effect. In this 
case the entropy $H\left(n,E\vert N,T,\lambda\right)$ does not depend 
on $\left\{\lambda\right\}$, and expression \eqref{H(lambda,n,E|N,T)}
is reduced to the sum
\begin{equation}
\ds
\label{H(lambda,n,E|N,T) simple}
	H\left(\lambda,n,E|N,T\right)
		= H\left(\lambda\right)
		+ H\left(n,E\vert N,T\right).
\end{equation}

\section{Conclusions}
\label{sec:conclusions}

The Unruh effect is considered from the point of view of information 
theory. We estimated the total entropy of the radiation generated by 
the Unruh horizon in the non-inertial reference frame for the state 
verified as vacuum by any inertial observer. Usually such a case is 
treated as von Neumann entropy of the corresponding density matrix.
But this is just the starting point of our study, because the density 
matrix of outgoing radiation describes conditional multiplicity 
distribution at given energy and Unruh temperature. As a result, it 
allows one to estimate \emph{total} entropy of the Unruh source by
taking into account both multiplicity and energy distribution of the 
outgoing quanta. We show how it can be calculated even without exact 
knowledge of the corresponding Hamiltonian. In particular, such a lack 
of information results in Schwinger-like spectrum of the emission, 
see Eq.~\eqref{p(E)}.

The case of finite amount of particle emission is considered. It allows 
us to utilize the results for realistic particle emission spectra.
Asymptotics of the general expression for entropy with respect to low
and high values of Unruh temperature are also investigated.
We found that in case of small acceleration, corresponding to low 
temperature, entropy of the radiation does not depend on maximal amount 
of emitted particles in the leading order, 
see Eq.~\eqref{H(n,E|N,T) at T to 0 case B}.
Dependence on $N$ is recovered for large accelerations, when 
$ T \to \infty $, see Eq.~\eqref{H(n,E|N,T) at T to infty case B}. It 
can be explained by abundant emission of particles from the hot Unruh 
horizon, when amount of the emitted quanta may be considered as extra 
degree of freedom contributing to the total entropy.

Another interesting point is that the total entropy 
$ H\left(n,E|N,T\right) $ quickly drops to zero with increase of the 
mass $m$ of the quanta. It can be explained by the energy 
conservation law: the more energy is being spent on creation of 
particle's mass, the less of it may be used to generate total 
distribution. At the same time, total entropy of the Unruh source 
slightly increases with the maximum allowed energy $ M $, because the 
distribution widens with increase of $M$ thus leading to the total 
entropy increase.

The obtained results can be applied to analysis of particle 
distributions in inelastic scattering processes at high energies. Also, 
they may be generalized to other degrees of freedom of the emitted 
particles, such as spin, charges etc. However, such a generalization 
may significantly complicate the analysis. For instance, additional 
conservation laws originating from the other degrees of freedom might 
change the metric. Therefore, one may be forced to take distribution 
$ \left\{T\right\} $ into account too.

\begin{acknowledgments}
Fruitful discussions with J.M.~Leinaas, O.~Teryaev and S.~Vilchinskii 
are gratefully acknowledged.
M.T. acknowledges financial support of the Norwegian Centre for 
International Cooperation in Education (SIU) under grant 
``CPEA-LT-2016/10094 - From Strong Interacting Matter to Dark Matter".
The work of L.B. and E.Z. was supported 
by the Norwegian Research Council (NFR) under grant No. 255253/F50 -
``CERN Heavy Ion Theory" and 
by Russian Foundation for Basic
Research (RFBR) under Grant No. 18-02-40084 and Grant No. 18-02-40085.
Numerical calculations and visualization were made at Govorun (JINR, 
Dubna) computer cluster facility.
\end{acknowledgments}

\end{document}